\documentclass[12pt]{article}
\usepackage{amsmath,amsfonts,amssymb,amscd,color}
\usepackage{amstext}
\usepackage{latexsym}
\usepackage{amssymb}

\def\R{\mathbb R}

\def\N{\mathbb N}

\newcommand{\eps}{\varepsilon}

\begin{document}

\newtheorem{theorem}{Theorem}[section]
\renewcommand{\thetheorem}{\arabic{section}.\arabic{theorem}}
\newtheorem{definition}[theorem]{Definition}
\newtheorem{deflem}[theorem]{Definition and Lemma}
\newtheorem{lemma}[theorem]{Lemma}
\newtheorem{example}[theorem]{Example}
\newtheorem{remark}[theorem]{Remark}
\newtheorem{remarks}[theorem]{Remarks}
\newtheorem{cor}[theorem]{Corollary}
\newtheorem{pro}[theorem]{Proposition}
\newtheorem{proposition}[theorem]{Proposition}

\renewcommand{\theequation}{\thesection.\arabic{equation}}

\title{Comments on the paper `Static solutions 
of the Vlasov-Einstein system' by G.~Wolansky} 

\author{H{\aa}kan Andr\'{e}asson\\
        Mathematical Sciences\\
        Chalmers University of Technology\\
        University of Gothenburg\\
        S-41296 G\"oteborg, Sweden\\
        email: hand@chalmers.se\\
        \ \\
        Markus Kunze\\
        Mathematisches Institut\\
        Universit\"at K\"oln\\
        Weyertal 86-90\\
        D-50931 K\"oln, Germany\\
        email: mkunze@mi.uni-koeln.de}

\maketitle

\begin{abstract}
\noindent
In this note we address the attempted proof of the existence of static solutions 
to the Einstein-Vlasov system as given in \cite{Wol}. We focus on a specific and central part 
of the proof which concerns a variational problem with an obstacle. We show that two important claims 
in \cite{Wol} are incorrect and we question the validity of a third claim. We also discuss 
the variational problem and its difficulties with the aim to stimulate further investigations 
of this intriguing problem: to answer the question whether or not static solutions 
of the Einstein-Vlasov system can be found as local minimizers of an energy-Casimir functional. 

\end{abstract}


\section{Introduction}

\setcounter{equation}{0}
Static solutions of the Einstein-Vlasov system provide realistic models for galaxies \cite{BT} 
and it is an important problem to study the nonlinear stability of such solutions. 
Presently there is no rigorous proof of nonlinear stability for any non-trivial class of static solutions. 
One method to attack this problem is to show that static solutions can be obtained 
as minimizers of a variational problem. In the Newtonian case, i.e., for the Vlasov-Poisson system, 
this method has been successful, cf.~\cite{Rn}. In fact, in that case stability follows straightforwardly 
for solutions which have been obtained as minimizers. The situation is different for the Einstein-Vlasov system 
and stability does not follow directly from an analogous result. Nevertheless it would be very interesting 
to know if there are static solutions which are local minimizers of an energy-Casimir functional, 
and moreover, an affirmative result would provide a platform for attacking the nonlinear stability problem. 

In \cite{Wol}, Wolansky investigated this question and the asserted result is that there 
is indeed a class of static solutions to the Einstein-Vlasov system 
that can be constructed as local minimizers of an energy-Casimir functional. 
This work contains many interesting and innovative ideas but also statements which are incorrect 
or very difficult to justify rigorously. In this note we discuss a particular subproblem 
which shows up in \cite{Wol}: a variational problem with an obstacle and an additional derivative constraint. 
In order to solve this problem several difficulties must be overcome. We address three claims 
in the proof given in \cite{Wol} and we demonstrate that two of these claims are incorrect 
and we question the validity of a third one. 


Whether or not the main result in \cite{Wol} is true remains an open question. 
Several insights in \cite{Wol} indicate that this could be indeed the case. 
We hope that the present note stimulates further investigations of this intriguing problem. 

We have communicated the issues we raise in this work to the author of \cite{Wol} 
and he agreed that a note of this kind will be justified.

This paper is organized as follows. In the next section we formulate the subproblem mentioned above 
as it appears in \cite{Wol}. In section 3 we demonstrate that a central claim in \cite{Wol} is incorrect. 
In section 4 we question the claim that the minimizer solves the Euler-Lagrange equation, 
whereas section 5 contains a counterexample to an inequality stated in \cite{Wol}. Finally, 
in section 6 we briefly discuss a further technical difficulty associated 
with the variational problem formulated in section 2.    

\section{The variational problem with an obstacle}


In this note we focus on \cite[Section 5]{Wol}, which is key to this paper 
and where a variational problem is set up as follows. Consider the Lagrangian function 
\[ L(m, m', r)=\frac{r^{k+5/2}}{\sqrt{r-2m}}\,{\cal G}\Big(\frac{m'}{r^{2(k+1)}}\Big) \] 
from \cite[(32)]{Wol}, where ${\cal G}$ has the properties as described in \cite[Prop.~4.1]{Wol} 
and is defined in terms of a polytrope-type ansatz for the static solution; 
we shall take $k=0$ throughout to simplify the presentation. The associated action 
functional is 
\[ {\cal L}_R(m)=\int_0^R L(m(r), m'(r), r)\,dr \] 
which is to be minimized over the set of admissible functions 
\begin{eqnarray*} 
   \Gamma_R^M & = & \Big\{m: [0, \infty[\to [0, M]\,\,{\rm is\,\,absolutely\,\,continuous, 
   nondecreasing},
   \\ & & \hspace{1em} m(0)=0, m(r)=M\,\,{\rm for}\,\,r\ge R, 
   m(r)<Q(r)\,\,{\rm for}\,\,r\in ]0, R_1[\Big\},
\end{eqnarray*}  
where $Q(r)=(C_0/3)r^3-(c_0/4) r^4$ for $r\in [0, R_1]$ and $Q(r)=Q(R_1)$ 
for $r\ge R_1$ acts as a kind of barrier function, or obstacle, so that the term $\frac{2m(r)}{r}$ 
stays away from $1$ for $m\in\Gamma_R^M$. The numbers $R$, $M$, $R_1$, $C_0$ and $c_0$ 
are fixed; $M$ is the total mass of the system and later the limit $R\to\infty$ will be taken. 
The functional ${\cal L}_R$ is not bounded from below, so for $N\in\N$ large the Lagrangian $L$ 
is cut off and replaced by $L_N(m, m', r)=\frac{r^{5/2}}{\sqrt{r-2m}}
\,{\cal G}_N(\frac{m'}{r^{2k}})$ with ${\cal G}_N(s)=\max\{{\cal G}(s), s^2-N\}$. 
The corresponding action function is denoted by ${\cal L}_R^N$, it is bounded from below 
(when the norm of $W^{1, 1}$ is used) and admits a minimizer $m_N$ 
(whose the dependence on $R$ is omitted from the notation).   

In mathematical terms, one has to deal with an obstacle problem, the obstacle being $Q(r)$, 
under an additional derivative constraint, since functions in $\Gamma_R^M$ are nondecreasing. 
In the following three sections we are going to outline three serious issues 
with the arguments in \cite[Section 5]{Wol} related to the variational problem. 
\smallskip 

\noindent
{\bf Remark:} We focus on the subproblem as described above. However, 
we point out that also other parts in \cite{Wol} do contain statements 
whose details we have not been able to fill in.  


\section{Equation (54) on p.~227}

\setcounter{equation}{0}

In \cite[p. 227]{Wol} it is claimed that $m_N(r)\le Q(\tau, r)$ 
holds for $\tau>0$ small enough, where 
\[ Q(\tau, r)=(C_0/3)r^3-((c_0+\tau)/4) r^4 \] 
is below $Q(r)$. For this, \cite[(54)]{Wol} asserts that 
\begin{eqnarray}\label{54}
   \lefteqn{\frac{d}{d^+\tau}\,{\cal L}_R^N(m_{N,\tau})}
   \nonumber
   \\ & = & \int_{K_{\tau}}\Big(\frac{\partial L_N}{\partial q}(m'_{N,\tau}, m_{N,\tau}, r)
   -\frac{d}{dr}\frac{\partial L_N}{\partial p}(m'_{N,\tau}, m_{N,\tau}, r)\Big)
   \,\frac{\partial m_{N,\tau}}{\partial\tau}\,dr.\qquad
\end{eqnarray}
Here 
\[ m_{N,\tau}(r):=\min \{m_N(r),Q(\tau,r)\},\;\; r\in [0,\infty[, \]
and $K_{\tau}$ is the open subset of $[0,R_{\tau}]$ such that 
\[ m_{N,\tau}(r)=Q(\tau,r)<m_N(r),\;r\in K_\tau,\mbox{ and } m_{N,\tau}(r)=m_N(r), \;r\notin K_\tau. \]
The number $R_{\tau}$ is fixed such that 
\begin{equation}\label{condition}
   \frac{\partial L_N}{\partial q}(Q'(\tau, r), Q(\tau, r), r)
   -\frac{d}{dr}\frac{\partial L_N}{\partial p}(Q'(\tau, r), Q(\tau, r),r)>0
\end{equation}
for any $r\in [0, R_{\tau}[$ 
and the prime denotes $\frac{\partial}{\partial r}$. 
Moreover, $m_{N,\tau}$ is forward differentiable and 
\begin{eqnarray}
   & &\frac{d}{d^+\tau}\,m_{N,\tau}(r)<0,\;\forall r\in K_{\tau},
   \label{Ktauless} \\
   & &\frac{d}{d^+\tau}\,m_{N,\tau}(r)=0,\;\forall r\notin K_{\tau}.
   \nonumber
\end{eqnarray}
Note that if (\ref{54}) holds and if $K_{\tau}$ has positive measure 
then it follows from (\ref{condition})-(\ref{Ktauless}) that 
\[ \frac{d}{d^+\tau}\,{\cal L}_R^N(m_{N,\tau})<0, \] 
which implies that 
\[ {\cal L}_R^N(m_{N,\tau})<{\cal L}_R^N(m_{N}), \] 
which contradicts the fact that $m_N$ is a minimizer, 
since $m_{N, \tau}\in\Gamma_R^M$ for $\tau>0$ small. 
Hence $K_{\tau}$ has zero measure and $m_N(r)<Q(0,r)=Q(r)$. 
\smallskip

We will now construct an example which demonstrates that (\ref{54}) is incorrect. 
We have chosen an explicit example since we find it instructive, 
but also since we want to check that no unexpected cancellations occur. 
In the explicit example we have computed the different terms numerically 
to confirm that this is not the case. 

\smallskip

\noindent
{\bf Remark:} In the example below the function $m_N$ is not a minimizer. 
As we will see, the claim (\ref{54}) is incorrect in this case. 
Of course, there is in principle a possibility that (\ref{54}) holds when $m_N$ is a minimizer. 
As will be clear later this would require that terms, which seem to be unrelated, do cancel. 
In any case, the argument in \cite{Wol} seems not to use the minimizing property 
at all, and in our communication the author of \cite{Wol} has agreed that (\ref{54}) is incorrect. 
\smallskip

The notation that we use in our example is slightly simplified from the notation in \cite{Wol}; 
the parameter $t=\tau/4$, the function $m$ replaces $m_N$, and $m_{N,\tau}$ is denoted by $h$. 
Moreover, we take $C_0=3$ and $c_0=4$. 
\smallskip

The example we construct reads as follows. Let $Q(t,r)=r^3-(1+t)r^4$ 
and consider the interval $r\in [0,1/2]$. 
For $r\in[0,1/4]$ let
\[
m(r)=r^3-\frac{1}{2}r^{7/2}=:m_1(r),
\]
and for $r\in ]1/4,1/2]$ let
\[
m(r)=r^3-\Big(\frac{3}{4}+r\Big)r^4=:m_2(r).
\]
We note that $m(r)<Q(0,r)$ everywhere except for $r=1/4$ where $m(r)$ and $Q(0,r)$ are equal.
Define $h(t,r):=\min\{m(r),Q(t,r)\}$. Let the Lagrangian function $L=L(p,q,r)$ be given by
\[ L(p,q,r)=-\frac{r^2}{1-\frac{2q}{r}}\log\Big(1+\frac{p}{r^2}\Big). \] 
The Lagrangian ${\cal L}$ then takes the form
\[ {\cal L}(m):=\int_0^{1/2} L(m',m,r)\,dr
   =-\int_0^{1/2} \frac{r^2}{1-\frac{2m(r)}{r}}\log\Big(1+\frac{m'(r)}{r^2}\Big)\,dr. \] 
Note that $-\log(1+s)$ has the right properties to qualify as a candidate 
for ${\cal G}$ (with $\alpha=1$), cf. Proposition 4.1 in \cite{Wol}, 
so that $L$ has the properties as required in \cite{Wol} in the case $k=0$. 

Let $t>0$ be given with $t\ll 1$. The functions $m$ and $h$ are identical 
except on a small interval $[r_1(t),r_2(t)]$, defining 
\[ r_1(t)=\frac{1}{4(1+t)^2},\quad r_2(t)=\frac14+t. \]
Clearly, the interval $[r_1(t),r_2(t)]$ corresponds to the set $K_\tau$ in \cite{Wol}. 
Since $h(t, r)=Q(t, r)$ for $r\in [r_1(t), r_2(t)]$, $h(t, r)=m_1(r)$ for $r\in [0,r_1(t)]$ 
and $h(r, t)=m_2(r)$ for $r\in [r_2(t),1/2],$ we obtain

\begin{eqnarray*}
   {\cal L}(h(t,\cdot))
   &=&\int_{0}^{r_1(t)}L(m_1'(r),m_1(r),r)\, dr+\int_{r_1(t)}^{r_2(t)}L(Q'(t,r),Q(t,r),r)\, dr
   \\ &&+\int_{r_2(t)}^{1/2}L(m_2'(r),m_2(r),r)\, dr. 
\end{eqnarray*}
Hence,
\begin{eqnarray*}
   \frac{d}{dt}\,{\cal L}(h(t,\cdot))&=&\dot{r}_1(t)L(m_1'(r_1),m_1(r_1),r_1)+\dot{r}_2(t)L(Q'(t,r_2),Q(t,r_2),r_2)
   \\
   && -\,\dot{r}_1(t)L(Q'(t,r_1),Q(t,r_1),r_1)-\dot{r}_2(t)L(m_2'(r_2),m_2(r_2),r_2)\\
   &&+\int_{r_1(t)}^{r_2(t)}\frac{\partial}{\partial t}\big(L(Q'(t,r),Q(t,r),r)\big) \, dr=:\sum_{k=1}^5 J_k.
\end{eqnarray*}
Here $\dot{r}_j=\frac{d}{dt}r_j$ and $r_j=r_j(t)$, $j=1,2$. 
In order to obtain a formula that is comparable to (\ref{54}) 
we rewrite the last term $J_5$ by integrating by parts. This yields
\begin{eqnarray*} 
   J_5& = &\int_{r_1(t)}^{r_2(t)}\bigg[\frac{\partial L}{\partial q}(Q'(t,r),Q(t,r),r)
   \frac{\partial Q}{\partial t}
   +\frac{\partial L}{\partial p}(Q'(t,r),Q(t,r),r)\frac{\partial Q'}{\partial t}\bigg]\,dr
   \\ & = & \int_{r_1(t)}^{r_2(t)}\bigg[\frac{\partial L}{\partial q}(Q'(t,r),Q(t,r),r)
   -\frac{d}{dr}\frac{\partial L}{\partial p}  (Q'(t,r),Q(t,r),r)\bigg]\frac{\partial Q}{\partial t} \, dr\\
   && +\,\frac{\partial L}{\partial p}(Q'(t,r_2),Q(t,r_2),r_2)\frac{\partial Q}{\partial t}(t,r_2)\\
   && -\,\frac{\partial L}{\partial p}(Q'(t,r_1),Q(t,r_1),r_1)\frac{\partial Q}{\partial t}(t,r_1).
\end{eqnarray*}
Altogether, we thus have
\begin{eqnarray*}
   \frac{d}{dt}\,{\cal L}(h(t, \cdot)) 
   &=&\dot{r}_1(t)L(m_1'(r_1),m_1(r_1),r_1)+\dot{r}_2(t)L(Q'(t,r_2),Q(t,r_2),r_2)
   \\
   && -\,\dot{r}_1(t)L(Q'(t,r_1),Q(t,r_1),r_1)-\dot{r}_2(t)L(m_2'(r_2),m_2(r_2),r_2)\\
   &&+\int_{r_1(t)}^{r_2(t)}\bigg[\frac{\partial L}{\partial q}(Q',Q,r)
   -\frac{d}{dr}\frac{\partial L}{\partial p}(Q',Q,r)\bigg]\frac{\partial Q}{\partial t} \, dr\\
   && +\,\frac{\partial L}{\partial p}(Q'(t,r_2),Q(t,r_2),r_2)
   \frac{\partial Q}{\partial t}(t,r_2)\\
   && -\,\frac{\partial L}{\partial p}(Q'(t,r_1),Q(t,r_1),r_1)
   \frac{\partial Q}{\partial t}(t,r_1)=:\sum_{j=1}^7 T_j 
\end{eqnarray*}
for $Q=Q(t, r)$. This expression should be compared to (\ref{54}) and we find that the integral term $T_5$ 
agrees with the right hand side in equation (\ref{54}), whereas there are in addition six boundary terms 
and there is no reason that they should cancel. Indeed, it is a straightforward task 
to numerically compute the terms $T_j$, $j=1, \ldots, 7$, for this explicit example 
and the boundary terms do in fact not cancel, since they have different orders of magintude. 
Hence equation (\ref{54}) does not hold for this example.

\section{Validity of the Euler-Lagrange equation}

\setcounter{equation}{0}

Suppose that by some means one would be able to show that $m_N(r)\le Q(\tau, r)$ 
for $\tau>0$ small enough. Then, on \cite[p.~227]{Wol} it is claimed (`As a result, we find ...') 
that as a consequence the Euler-Lagrange equation $\frac{d}{dr}\frac{\partial L_N}{\partial m'}
=\frac{\partial L_N}{\partial m}$ will follow. So let us fix some interval $[a, b]$ 
such that $a>0$ and a test function $\varphi$ which satisfies ${\|\varphi\|}_\infty\le 1$ 
and whose support is contained in $[a, b]$. Since $m_N\le Q(\tau, \cdot)$, 
certainly $m_N+\eps\varphi<Q(r)$ if $\eps$ is small enough. 
However, the set $\Gamma_R^M$ does have another boundary, induced by the condition 
that $m\in\Gamma_R^M$ should be increasing. Therefore, since one does not know 
that e.g.~$m'_N(r)\ge\delta>0$ for $r\in [a, b]$, it is not clear 
how good trial functions $m_N+\eps\varphi\in\Gamma_R^M$ could be obtained; 
rather than the Euler-Lagrange equation, one could only derive an inequality. 


\section{Last line on p.~227}

\setcounter{equation}{0}

In order to take the limit $N\to\infty$ on \cite[p.~228]{Wol}, 
it is required for the minimizer $m_N=m_N(r)$ that $\sup_{r\in ]0, R]} r^{-2} m'_N(r)\le C$ 
is finite, and for that it would be needed that $\lim_{r\to 0^+} r^{-2} m'_N(r)\le C$ is finite, 
see the last line on p.~227 and the beginning of p.~228. 

However, this fact cannot in general be deduced from $m_N$ being 
an absolutely continuous and non-decreasing function which satisfies  
$m_N(0)=0$ and $m_N(r)\le Q(r)=(C_0/3)r^3-(c_0/4)r^4$ for $r\in [0, R_1]$. 
To have an example, let 	$C_0=6$ and $c_0=4$, so that $Q(r)=2r^3-r^4$; 
the general case is analogous. Then $Q$ is increasing on $[0, 1]$ 
and such that $Q(1)=1$. Let $r_k=2^{-k}$ and $s_k=r_k-2^{-2k-2}\in ]r_{k+1}, r_k[$ 
for $k\in\N_0$. In addition, define $m: [0, 1]\to\R$ by $m(0)=0$, 
\[ m(r)=\left\{\begin{array}{c@{\quad:\quad}c}
   Q(r_{k+1}) & r\in [r_{k+1}, s_k], \\[1ex]
   \frac{Q(r_k)-Q(r_{k+1})}{r_k-s_k}\,(r-s_k)+Q(r_{k+1}) & r\in [s_k, r_k]. 
   \end{array}\right. \]
Then $m$ is continuous, non-decreasing and such that $m(r)\le Q(r)$ for $r\in [0, 1]$. 
Its (non-negative) derivative is given by 
\[ m'(r)=0,\,\,r\in ]r_{k+1}, s_k],
   \quad\mbox{and}\quad m'(r)=\frac{Q(r_k)-Q(r_{k+1})}{r_k-s_k},
   \,\,r\in [s_k, r_k]. \] 
In particular, $m'\in L^1(]0, 1[)$, since 
\begin{eqnarray*} 
   \int_0^1 m'(r)\,dr & = & \sum_{k=0}^\infty\int_{s_k}^{r_k} m'(r)\,dr
   =\sum_{k=0}^\infty(Q(r_k)-Q(r_{k+1})) 
   \\ & = & Q(r_0)-\lim_{n\to\infty} Q(r_{n+1})=1.  
\end{eqnarray*} 
This shows that $m\in W^{1, 1}(]0, 1[)$, whence $m$ is absolutely continuous. 
On the other hand, if $r\in ]s_k, r_k[$, then 
\begin{eqnarray*} 
   \frac{m'(r)}{r^2} & \ge & \frac{m'(r)}{r_k^2}=\frac{Q(r_k)-Q(r_{k+1})}{r_k^2(r_k-s_k)}
   =\frac{2r_k^3-r_k^4-2r_{k+1}^3+r_{k+1}^4}{2^{-2k-2}\,r_k^2}
   \\ & = & 2^{k+2}\,\Big(\frac{3}{4}-\frac{15}{16}\,2^{-k}\Big). 
\end{eqnarray*} 
Therefore in fact $\sup_{r\in ]0, 1]} r^{-2} m'(r)=\infty$,  
and whenever $(\xi_k)$ is a sequence such that $\xi_k\in ]s_k, r_k[$, 
then $\lim_{k\to\infty}\xi_k^{-2} m'(\xi_k)=\infty$ diverges.  


\section{A further technical difficulty}

Our main goal has been to obtain a rigorous proof of the claimed main result in \cite{Wol}. 
We thus have tried to show that either the statements in \cite{Wol} do hold, or to find other arguments 
that lead to the conclusion that there is a class of static solutions which are local minimizers 
of the energy-Casimir functional given in \cite{Wol}. 

An important step in understanding the variational problem formulated in section 2 
is to show that the local minimizer $m_N$ stays strictly away from the boundary, 
i.e., from the obstacle $Q(r)=(C_0/3)r^3-(c_0/4) r^4$. 
As in \cite{Wol}, it is natural to try to show this claim by a contradiction argument. 
Let us denote by $K=\{r: m_N(r)=Q(r)\}$ the coincidence set. If we assume that $K$ is an interval, 
then we are able to obtain the desired contradiction and it would follow that $m_N<Q$. 
However, $K$ can be a very complicated set and in this case we did not succeed 
in completing the argument. Of course, since the claim holds for $K$ being an interval, 
this could be an indication that the claim is in fact true.

\end{document}